\documentclass[aps,pra,twocolumn,showpacs,preprintnumbers,amsmath,amssymb]{revtex4}

\usepackage{graphicx}
\usepackage{dcolumn}
\usepackage{bm}

\begin{document}
\title{Geometry- and diffraction-independent ionization probabilities in intense laser
fields:\\probing atomic ionization mechanisms with effective
intensity matching}
\author{W. A. Bryan}
 \email{w.bryan@ucl.ac.uk}
\author{S. L. Stebbings}
\author{E. M. L. English}
\author{T. R. J. Goodworth}
\author{W. R. Newell}
 \email{w.r.newell@ucl.ac.uk}
\affiliation{Department of Physics and Astronomy, University
College London, Gower Street, London WC1E 6BT, UK}
\author{J. McKenna}
\author{M. Suresh}
\author{B. Srigengan}
\author{I. D. Williams}
\affiliation{Department of Pure and Applied Physics, Queen's
University Belfast, Belfast BT7 1NN, UK}
\author{I. C. E. Turcu}
\author{J. M. Smith}
\author{E. J. Divall}
\author{C. J. Hooker}
\author{A. J. Langley}
\affiliation{Central Laser Facility, Rutherford Appleton
Laboratory, Chilton, Didcot, Oxon OX11 0QX, UK}
\date{\today}

\begin{abstract}
We report a novel experimental technique for the comparison of
ionization processes in ultrafast laser pulses irrespective of
pulse ellipticity. Multiple ionization of xenon by 50 fs 790 nm,
linearly and circularly polarized laser pulses is observed over
the intensity range 10 TW/cm$^2$ to 10 PW/cm$^2$ using Effective
Intensity Matching (EIM), which is coupled with Intensity
Selective Scanning (ISS) to recover the geometry-independent
probability of ionization. Such measurements, made possible by
quantifying diffraction effects in the laser focus, are compared
directly to theoretical predictions of multiphoton, tunnel and
field ionization, and a remarkable agreement demonstrated. EIM-ISS
allows the straightforward quantification of the probability of
recollision ionization in a linearly polarized laser pulse.
Furthermore, probability of ionization is discussed in terms of
the Keldysh adiabaticity parameter $\gamma$, and the influence of
the precursor ionic states present in recollision ionization is
observed for the first time.
\end{abstract}

\maketitle

\section{Introduction}
Over the last decade and a half, the advent of modern laser
technology has resulted in a fruition of significant experimental
studies on the interaction of intense ultrafast laser pulses with
dilute matter. The success of these works has naturally
precipitated a reaction from the theoretical community, which has
lead to a highly productive dialogue. Particular highlights are
the observation \cite{hhgexp} of high harmonic generation (HHG),
the use of HHG to generate attosecond-timescale bursts of XUV
radiation \cite{atto}, nonsequential ionization of atoms
\cite{nonseqat} and molecules \cite{nonseqmol} through electron
rescattering and the enhanced ionization of molecules leading to
Coulomb explosion \cite{postrev}. Experimental advances by the
authors and co-workers have allowed the investigation of the
interaction of ultrafast laser pulses with ionic targets, where
the ionization of ground and metastable Ar$^{+}$ \cite{green},
C$^{+}$ \cite{steb} and Xe$^+$ \cite{suresh} has been observed for
the first time. Central to all of these studies is the observation
of process which are dependent on focused laser intensity,
traditionally facilitated by variation of the pulse energy. While
being straightforward to carry out experimentally, this technique
suffers the drawback that, as the pulse energy is varied, the size
of the laser focus is altered. Such geometry-dependent
measurements are of course still of great importance, and have
allowed major advances in the understanding of laser-matter
interactions: the drawback is in making comparisons with theory. A
different approach has been facilitated by the ongoing development
of laser amplification techniques \cite{backus}, permitting
macroscopic control of laser focal conditions through intensity
selective scanning (ISS) \cite{walk, hansch}. Here the laser focal
geometry is constant, and a spatially selective detector is used
to image different regions of the focus. Such measurements are of
particular interest to the atomic physics community, with regards,
for example, to the response of noble gas atoms to focused
femtosecond laser pulses.

While a number of experimental studies have compared atomic
ionization using circular and linear polarized laser pulses, for
recent examples see \cite{guo, nonseqmol}, to the authors'
knowledge, all have employed the traditional geometry-dependent
technique. For a particular pulse energy, there is a change in
electric field amplitude when the polarization is switched from
linear to circular, making a comparison problematic. We propose a
new method to circumvent this problem, as a development of ISS. A
constant pulse energy ratio between linear and circular
polarizations is defined such that the relative ionization yield
is constant as the laser focus is translated past the spatially
selective detector, irrespective of laser polarization. This
technique of effective intensity matching (EIM) coupled with ISS
is employed to examine ultrafast strong field ionization of xenon:
EIM-ISS results are presented.

Before quantitative analysis of the EIM-ISS results is possible,
it is necessary to remove the geometry dependence inherent in the
ionization yields. A method to perform such a deconvolution has
been proposed in \cite{walk}. An analogous technique has been
presented by the authors \cite{toby}, where a numerical inversion
is employed to remove the geometry dependence from the ISS
results. However both techniques only allow the removal of
Gaussian focusing. Modern laser systems typically generate pulses
with a Gaussian profile in the far-field. However, optical
transport systems impose geometrical constraints such that it is
practically impossible to propagate this perfect profile into the
experimental chamber. Such beam propagation has been addressed
recently by L\"{u} and co-workers \cite{luji} through the
application of an aperture function approximation \cite{wen} to a
multi-apertured optical system. Unfortunately, this approximation,
even with little or no beam truncation, will introduce oscillatory
distortions to the laser intensity. Zhang and co-workers
\cite{zhang} have presented an elegant analysis of the diffraction
of a focused beam, with particular application to traditional
geometry-dependent intensity variation measurements. In the
present work, beam diffraction is quantified before focusing takes
place: this treatment better representing typical experimental
systems.

The spatial insensitivity of most instruments employed in
intensity variation measurements \cite{kondo, shee, dimauro, tale,
laro, lafon, guo} average over such diffraction effects, as all of
the confocal volume lies within the volume that the instrument
images. In the case of EIM-ISS, the very spatial sensitivity that
makes the technique so powerful also makes it susceptible to
optical distortions. Given that the EIM-ISS technique naturally
discriminates between low-intensity large volume and
high-intensity small volume processes, then before the laser -
atom system under investigation can be fully understood, it is
vital to quantify the spatial distribution of laser intensity.
Herein, we present the results of an analytical treatment of the
focusing of a truncated Gaussian laser pulse through an arbitrary
ABCD (where A to D are elements of the System Matrix) optical
system, the derivation of which is given in Appendix A.

By matching the optical conditions in our ABCD model to those in
our experimental focus, we then remove the geometry dependence
from the EIM-ISS data, revealing for the first time geometry-free
and diffraction-free probabilities of ionization for both linear
and circular polarizations, and compare them directly to
theoretical predictions.

\section{Application to laser - dilute matter interactions}
In the following section, the optical system employed is defined,
then the xenon EIM-ISS data is presented. The apertured solution
for an arbitrary ABCD optical system is then employed to remove
the geometrical influence of the volume of the laser focus,
resulting in geometry-independent atomic ionization probabilities.
Ionization mechanisms are then discussed in detail. Finally, the
variation of geometry-independent ionization probability with the
Keldysh parameter \cite{keld} is presented, allowing ionization
mechanisms to be discussed in terms of the relative frequencies of
the tunneling electron and the laser field.

\subsection{Optical system}
\begin{figure}
\includegraphics[width=235pt]{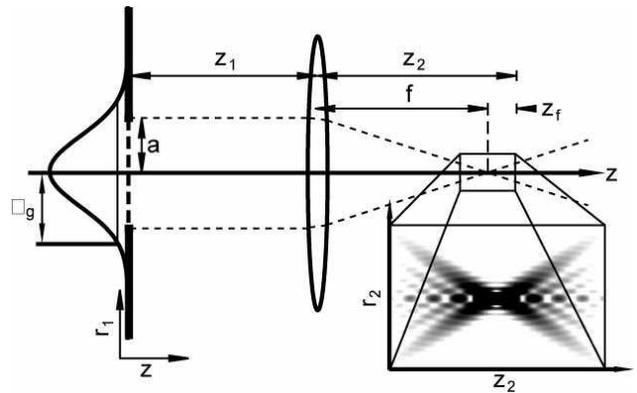}\\
\caption{Illustration of the cylindrically symmetric optical
system employed in recent laser - dilute matter interactions. The
aperture, radius {\it a}, lies in the input plane at {\it z} = 0,
with radial co-ordinate {\it r}$_1$. The resulting focus lies in
the output plane at {\it z} = {\it z}\hspace{1pt}$_1$ + {\it
z}\hspace{2pt}$_2$, with radial co-ordinate {\it r}$_2$. The focal
length of the lens is {\it f}, and {\it z$_f$} = {\it z}$_2$ -
{\it f}\hspace{2pt} is position with respect to the
focus.}\label{fig1}
\end{figure}

\begin{figure}
\includegraphics[width=235pt]{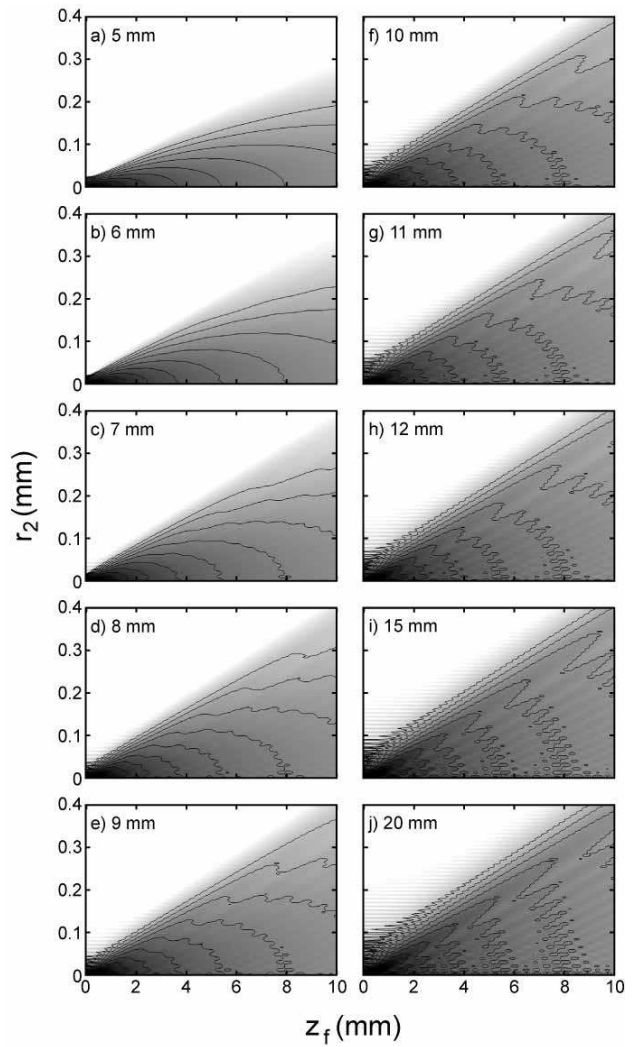}\\
\caption{(a) to (j) Simulated intensity distributions in
logarithmic grayscale with isointensity contours, separated by an
order of magnitude, for the optical system in Fig. 1. The aperture
radius {\it a} = 11 mm is kept constant, while the beam radius
$\omega_g$ is varied from 5 to 20 mm. Diffraction is clearly
apparent even when the aperture is considerably larger than the
beam radius.}\label{fig2}
\end{figure}

In recent experimental studies published by the authors and
co-workers \cite{green, steb, suresh, toby} the optical system is
as illustrated in Fig.~\ref{fig1}, with a corresponding system
matrix
\begin{equation}
\left(%
\begin{array}{cc}
  A & B \\
  C & D \\
\end{array}%
\right) = \left(%
\begin{array}{cc}
  1-(z_2/f)  &  \hspace{10pt}z_1+(1-z_1/f)z_2\\
  -1/f  &  1-(z_1/f)\\
\end{array}%
\right)
\end{equation}
where {\it z}$_1$ and {\it z}$_2$ correspond to the distance
between the aperture and lens, and lens and focus, and {\it f} is
the focal length of the lens. We also define {\it z}$_f$ = {\it
z}$_2$ - {\it f} as the position parallel to the beam propagation
direction with respect to the focus. The system matrix is derived
using straightforward matrix optics, and is the transformation
from the input plane (containing the aperture) to the output
plane, through a translation, a refraction and a second
translation to the focus. In the present work, the following
values apply: {\it z}$_1$ = 300 mm, {\it z}$_2$ = 250 mm, {\it f}
= 250 mm and the aperture radius {\it a} = 11 mm. Given the
wavelength, $\lambda$ = 790 nm, the only quantity which is unknown
is the beam radius $\omega_g$ before the aperture at which
diffraction occurs. The fixed aperture, located at {\it z} = 0,
defines the reference plane for the System Matrix.
As presented in Appendix A, we derive a solution for the laser
intensity distribution in the vicinity of the focus, which
accounts for diffraction of the incoming laser beam at an aperture
of finite diameter. Such a solution is particularly applicable to
ultrafast (Ti: sapphire based) physics, where the intensity
distribution in the laser focus is rarely Gaussian.
Fig.~\ref{fig2}(a)\hspace{1pt}-(j) shows the result of simulating
the focus of the current optical system for the fixed radius
aperture and a range of beam radii. Initially, the beam radius
$\omega_g$ = 5 mm, and the focus created is almost identical to
the unapertured form described by Eq. (A6) in Appendix A. However,
as the beam radius is increased, the focus is disturbed by the
diffraction of the incoming beam at the aperture. As the beam
radius becomes comparable to the aperture radius, the focus shows
pronounced lines of maximum and minimum intensity in both {\it
z}$_f$ and {\it r}$_2$. The isointensity contours in
Fig.~\ref{fig2}(a)\hspace{1pt}-(j), separated by an order of
magnitude, illustrate that as the beam radius is increased, the
focal spot size increases along {\it r}$_2$ contrary to the
behaviour expected in the case of an unapertured Gaussian beam.

\subsection{Effective intensity matching (EIM)\\as applied to ISS}
The experimental apparatus employed in the present work has been
described in detail \cite{green, steb, suresh, toby, wab}, so only
a brief outline is given here. The 20 mJ, 790 nm 50 fs output of
the ASTRA Ti:sapphire laser is transmission focused on to the
target gas in the source region of the time-of-flight mass
spectrometer. Given the high power of the laser used in our recent
work, intensities in excess of 10$^{17}$ Wcm$^{-2}$ are routinely
generated in a long focus. Indeed, the intensity range required
for ultrafast ionization rate studies, typically $>$ 10$^{13}$
Wcm$^{-2}$ is produced $\approx$ 10 mm from the centre of the
focus. The 250 $\mu$m entrance aperture located in the
time-of-flight mass spectrometer places a tight spatial limitation
on the volume of the focused laser beam which produces ion signal
from the target gas, and by translating the focusing lens parallel
to the z-axis, the spectrometer is exposed to different regions of
the focus. Typically, such ISS or {\it z}-scan experiments are
performed by averaging the ion signal measured at each lens
position, {\it z}$_f$.

To explore the ionization mechanism in xenon we have developed a
novel experimental technique, namely effective intensity matching,
combined with intensity selective scanning (EIM-ISS). The essence
of EIM-ISS is to define a constant ratio, {\it R}$_{\rm EIM}$
between the laser intensities of the linearly ({\it
I}\hspace{2pt}$_{\rm lin}$) and circularly ({\it
I}\hspace{2pt}$_{\rm circ}$) polarized laser beams such that the
spatial distribution of the ions detected for each polarization
are the same for all {\it z}$_f$ values assuming that
nonsequential ionization processes are negligible. Before
recording the xenon data presented, the ionization of neon was
observed with circularly and linearly polarized radiation to
define the ratio {\it R}$_{\rm EIM}$. Neon is a good test gas as,
of all the noble gases, it is least susceptible to nonsequential
(recollision) ionization \cite{neonrecol}. Our studies, not
presented here \cite{suresh}, reveal that {\it R}$_{\rm EIM}$ =
{\it I}\hspace{2pt}$_{\rm lin}$ / {\it I}\hspace{2pt}$_{\rm circ}$
= 0.65 $\pm$ 0.02 gives a remarkable match in the Ne$^+$
ionization signal over a large ($\approx$ 10 mm) range of {\it
z}$_f$, equivalent to an intensity range 10$^{13}$ Wcm$^{-2}$ to
10$^{16}$ Wcm$^{-2}$. A similar approach was reported in
\cite{auguste} to explain threshold ionization intensities
observed with linearly and circularly polarized light in a
long-pulse (1 ps) traditional intensity variation measurement.
However, the present study and that of \cite{suresh} are of a
higher precision, being the culmination of a systematic
investigation, and are unique in the area of ultrafast intense
field interactions.

The exact value of {\it R}$_{\rm EIM}$ is determined by the
influence of the different laser polarizations and electric field
amplitudes on the ionization process. In the linearly polarized
case, the sinusoidal laser electric field is modulated by the
pulse envelope: the field amplitude oscillates under a typically
sech$^2$ temporal profile. With circular polarization the electric
field is continually present, and the electric field direction
rotates through 2$\pi$ during the laser period, and the field
amplitude typically takes a sech$^2$ profile. In the circularly
polarized field, the projection of the angular momentum imparted
to the electron is conserved along the direction of beam
propagation. However, in the linearly polarized field, momentum is
imparted to the electron in the direction of electric field
oscillation, i.e. perpendicular to the propagation axis. This has
an important influence on tunnel ionization, as recently noted by
Tulenko and Zon \cite{zoncirc} in which the rate of tunnelling
depends on the magnetic quantum number of the tunnelling electron:
in \cite{ppt} it was concluded that the ionization rate should be
identical, irrespective of polarization type.

\begin{figure}
\includegraphics[width=235pt]{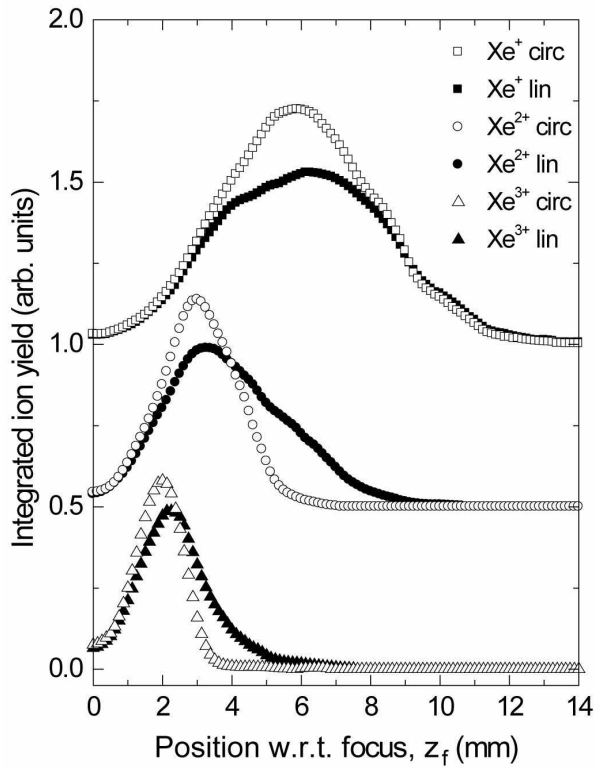}\\
\caption{Raw EIM-ISS data for the ionization of xenon to Xe$^{n+}$
({\it n} = 1, 2, 3). The integrated xenon ion yield is recorded as
function of focusing optic position with respect to the axis of
the spectrometer, in Fig. 1 equivilant to {\it z}$_f$. Linear
(solid symbols) and circular (open symbols) polarized laser pulses
are employed, where EIM is used to \emph{define} the distribution
of ion signal with varying {\it z}$_f$. The presence of
nonsequential (recollision) ionization is apparent in the
Xe$^{2+}$ and Xe$^{3+}$ signal as signal enhancement at low
intensity (large {\it z}$_f$) in the case of linear polarization.
Ripples in the ion yield indicate the presence of
diffraction.}\label{fig3}
\end{figure}
With {\it R}$_{\rm EIM}$ defined, the spectrometer source region
was filled with xenon to a pressure of 1 $\times$ 10$^{-8}$ mbar:
a low pressure is used to avoid the influence of space-charge
effects. All ions generated were averaged over 500 laser shots for
each position along {\it z}$_f$. The observed {\it z}-scan is
presented in Fig.~\ref{fig3}, showing Xe$^{\it n+}$ ({\it n} = 1,
2, 3) product ion yields, with the Xe$^{2+}$ and Xe$^+$ yields
displaced vertically by 0.5 and 1.0 respectively for clarity. By
limiting the field of view of the spectrometer, the ion signal
from a particular charge state at each z-position is an integral
over all {\it r}$_2$ and a narrow range of {\it z}$_f$. This
selectivity is apparent in the data in Fig.~\ref{fig3}. Starting
at the focus, each ion yield presented is observed to increase
from near zero to a maximum, the position of the maximum moving to
lower {\it z}$_f$ with increasing charge state, {\it n}. Each
successive charge state requires an increase in intensity to
maximize its ionization yield, a common result of traditional
intensity measurements \cite{kondo, shee, dimauro, tale, laro,
lafon, guo}. Subtle ripples present in the data presented in
Fig.~\ref{fig3} are the result of diffraction in the ultrafast
laser focus. Importantly, linear and circular time-of-flight
spectra are recorded at each {\it z}$_f$ position as the focusing
optic is translated, rather than recording a linear scan followed
by a circular scan. This ensures that conditions within both the
spectrometer and laser focus undergo minimal variation.

Xenon has a far greater propensity for ionization through
recollision in a linear laser field than neon, giving rise to the
major differences between linearly and circularly polarized light
observed in the EIM-ISS data in Fig.~\ref{fig3}. The data recorded
with circular polarization is $\emph{recollision-free}$, as the
probability of the first-ionized electron returning to the ion in
a circularly polarized field is negligible \cite{cork}. The
`signature' of recollision is clearly apparent in the Xe$^{2+}$
and Xe$^{3+}$ data as an enhancement of the integrated ion yield
with linear polarization at positions of {\it z} at larger
distances from the focus than the maxima (for example, in
Xe$^{2+}$ 4.5 mm $<$ {\it z}$_f$ $<$ 8 mm). This enhancement is
due to recollision ionization in the linear field being more
efficient than sequential multiple ionization by the laser field
at low intensities (i.e. large {\it z}$_f$ with respect to the
centre of the focus).

The enhancement of ion yield by recollision ionization in
Xe$^{2+}$ and Xe$^{3+}$ is directly responsible for the
significant differences in the Xe$^+$ yield curves. This is a
consequence of the conservation of confocal volume, necessitated
by the following important points. The core result of EIM-ISS is
that the \emph{effective} laser intensity is identical for linear
and circular polarizations. Given that the number density is
constant, the ionization yield from sequential ionization of the
atoms in the confocal volume is made to match for both
polarizations over all {\it z}$_f$. However, the volume of the
laser focus must be conserved, therefore, we observe for the first
time the true interplay between the volumes generating a charge
state Xe$^{\it n+}$ ({\it n} = 1, 2, 3...), which totally contains
and often overlaps with the volume generating all higher charge
states. Therefore the suppression of integrated ion yield in the
case of Xe$^+$ in the range 3 mm $<$ {\it z}$_f$ $<$ 8 mm is due
to the depopulation of the Xe$^+$ volume in the laser focus (when
the laser is linearly polarized) through the observed mechanism of
recollision ionization to higher charge states. An analogous
depletion is observed around the maximum of the Xe$^{2+}$ yield,
caused by depopulation of the Xe$^{2+}$ volume by recollision
ionization to Xe$^{3+}$ in linearly polarized radiation. This
sequence is repeated for the higher charge states. By comparing
the sum of ion yields over all charge states at each {\it z} for
the two polarizations, a near exact match is observed, confirming
conservation of focal volume. This then is the major benefit of
EIM-ISS, allowing $\textit{direct}$ comparisons to be made between
atomic ionization mechanisms for the first time.

\subsection{Deconvolution of EIM-ISS results:
volume-independent ionization probabilities} To further develop
our understanding of the interplay of ionization mechanisms as
observed using EIM-ISS using both linear and circular
polarizations, we turn now to removing the geometry dependence of
the ionization signal from the EIM-ISS data in Fig.~\ref{fig3}.
The aim of such analysis is the recovery of geometry-independent
ionization probabilities.

Van Woerkom and co-workers have previously established a technique
for the removal of the dependence of ionization probability on
focal volume \cite{walk,hansch}, where the ionization probability
{\it $\Omega$}({\it I}\hspace{2pt}), may be calculated from the
{\it z}$_f$ dependent ion signal {\it S}({\it z}$_f$\hspace{1pt})
via the on-axis intensity distribution {\it I$_{ax}$}({\it
z}$_f$\hspace{1pt}) according to:
\begin{equation}
\Omega(I_{ax}(z_f)) \propto
\left(\frac{I_{ax}(z_f)}{dI_{ax}(z_f)/dz_f}\right)
\frac{d}{dz_f}\hspace{2pt}[I_{ax}(z_f)S(z_f)]
\end{equation}
In the present work, we have refined this treatment, firstly to
allow for the non-Gaussian nature of the focussed beam and
secondly removing the dependence of ionization probability on the
intensity distribution within the laser focus, despite the
presence of diffraction. The latter is achieved by applying Eq.
(A9) in Appendix A to Eq. (2) where {\it I$_{ax}$} = $|${\it
U}\hspace{2pt}({\it r}$_2$ = 0, {\it z}$_f$)$^2|$.
\begin{figure}
\includegraphics[width=235pt]{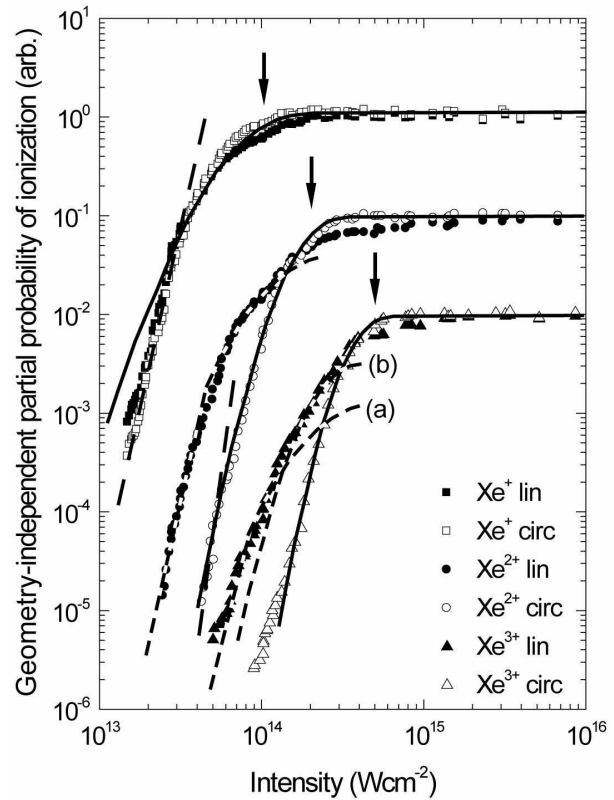}\\
\caption{EIM geometry-independent partial probability of
ionization (PPI) for Xe$^{\it n+}$ ({\it n} = 1, 2, 3) as function
of intensity, measured for both linear (solid symbols) and
circular (open symbols) polarizations. Increasing charge states
are normalized to descending orders of magnitude for clarity.
Three types of sequential ionization are identified: multiphoton
(low intensity, constant gradient, long dashed line), tunnelling
(intermediate intensity, decreasing gradient, solid line) and
field ionization (high intensity, vertical arrow). Recollision
ionization (short dashed line) is also observed with linear
polarization. See text for discussion.}\label{fig4}
\end{figure}

Although the 1/e$^2$ beam radius is difficult to define
accurately, it is of a comparable size to the final aperture
radius. Furthermore, the ionization signal of the first three
charge states is expected to saturate within the range of {\it
z}$_f$ (i.e. intensity) in the current experiment \cite{laro},
thus the gradient of the ionization probability should tend to
zero at high intensity for all change states, {\it n}. At
intensities above saturation, it is reasonable to expect
ionization probability to decrease, as ionization to higher charge
states becomes increasingly more likely. However, if we define
some saturation intensity {\it I}\hspace{2pt}$_{\rm SAT}$, which
occurs at a corresponding on-axis distance {\it
z}\hspace{2pt}$_{\rm SAT}$ from the focus, at {\it z}$_f$ $<$ {\it
z}\hspace{2pt}$_{\rm SAT}$ (i.e. on-axis {\it I} $>$ {\it
I}\hspace{2pt}$_{\rm SAT}$), {\it I}\hspace{2pt}$_{\rm SAT}$ is
still present off-axis thus occupying a larger proportion of the
volume than the on-axis contribution. The result of applying Eq.
(2) is only valid at intensities $<$ {\it I}\hspace{2pt}$_{\rm
SAT}$: at intensities $>$ {\it I}\hspace{2pt}$_{\rm SAT}$, we {\it
define} the probability of ionization as unity. This volume- and
diffraction-free intensity dependent quantity is referred to as
the partial probability of ionization (PPI). If the conserved
probability of ionization (CPI) is required, this may be readily
calculated according to CPI ({\it n}) = PPI ({\it n}) - PPI ({\it
n} + 1), where {\it n} is charge state, as before. This technique
will be applied in future publications.

The geometry- and diffraction-independent PPIs are now recovered
from the EIM-ISS data in Fig.~\ref{fig3}, and are presented in
Fig.~\ref{fig4}, where the saturated (unity) ionization
probabilities of the increasing charge states are normalized to
descending orders of magnitude to aid visual presentation. The
beam radius $\omega_g$ is estimated from a measurement of the
unfocused beam profile, and small adjustments (of the order of 0.1
mm) made to $\omega_g$ until the PPI curves recovered do not
exhibit rapid changes of gradient with intensity. The beam radius
$\omega_g$ = 10.25 mm used to recover these ionization
probabilities is found to be not only consistent over all charge
states observed in xenon ({\it n} = 1, 2, 3), but also for a wide
range of atomic (the other Noble gases) and molecular (H$_2$,
D$_2$, N$_2$, CO$_2$) targets covered in our studies.

To quantify the results presented in Fig. 4, the
intensity-dependent probability of ionization to Xe$^{\it n+}$
({\it n} = 1, 2, 3) by four ionization mechanisms is now
discussed. Starting with the lowest intensity, the following
mechanisms are apparent in Fig.~\ref{fig4}:

(i) Multiphoton ionization (MPI) is predicted by lowest-order
perturbation theory (LOPT) to vary according to $\it I^N$
\cite{mpi}, where {\it N} is the number of 790 nm (1.56 eV)
photons absorbed, thus on a log-log plot of probability vs.
intensity a constant gradient of {\it N} should exist. Increasing
charge state {\it n} requires an increasing number of photons. For
example, Xe$^{2+}$ production requires an energy of 21.21 eV,
therefore the absorption of at least fourteen 790 nm photons is
required. The expected gradients for MPI of Xe$^{n+}$ are 7.8
({\it n} = 1), 13.6 ({\it n} = 2) and 20.6 ({\it n} = 3). For {\it
n} = 1 and 2, the expected gradients are shown in Fig. 4 as long
dashed lines. For {\it n} = 3, there is no MPI observed within the
sensitivity of the experiment, determined by the low target gas
density.

(ii) Tunnel ionization (TI), as described initially by Keldysh
\cite{keld} and later refined by Popov and co-workers \cite{ppt},
is the result of a bound electron tunnelling out of the atom
through the laser-modified Coulomb potential. With increasing
intensity, TI becomes more efficient, as the laser + Coulomb
barrier width decreases. Represented in Fig. 4 as solid lines,
Keldysh theory predicts a tunnel rate, which we have converted to
probability. Note that the amount of TI observed is not always
well-predicted by tunnelling theory, as observed recently by
Yamakawa and co-workers \cite{yama} who introduced scaling factors
in order to match theory and experiment. In the present work, when
the TI theory is shifted to improve the fit with the PPI, the
factors used are made clear.

(iii) At the highest intensities, classical field ionization (FI)
dominates, as the electric field of the laser pulse is sufficient
to rapidly suppress the Coulomb potential, allowing direct
liberation of the electron. The intensity at which FI is predicted
to occur is indicated by the vertical arrows in Fig. 4, calculated
using the over-the-barrier model. The intensity at which FI
dominates should coincide with the saturation of TI.

Ionization mechanisms (i) to (iii) are sequential in nature, and
due to the EIM technique should occur at the same effective laser
intensity for linear and circular polarizations (evident from the
overlap of the Xe$^+$ PPIs in Fig. 4). As the ionization potential
increases with charge state, we expect a systematic increase in
the intensity required to produce a certain ionization mechanism.
Furthermore, as discussed with reference to Fig. 3, there is an
enhancement of signal in a linearly polarized laser pulse at low
intensities, thus we must also consider nonsequential ionization
processes:

(iv) Recollision ionization (RI) \cite{cork}, also referred to as
nonsequential double or multiple ionization (NSDI or NSMI),
predominantly occurs due to the linear laser field driving a
correlated liberated electron back to the parent ion, thus
initiating secondary ionization \cite{beckmos}. This process has
been the subject of a number of COLTRIMS studies \cite{coltrev}.
In Fig. 4, the enhancement of Xe$^{2+}$ and Xe$^{3+}$ by RI is
clearly seen in the difference of PPI curves for linear and
circularly polarized radiation. In Fig. 4, the presence of RI is
indicated by the short dashed lines, generated by transferring the
theoretical prediction of sequential ionization of previous charge
state(s) onto the linearly polarized data. For Xe$^{2+}$ RI, the
combined MPI and TI to Xe$^+$ PPI is fitted, whereas for Xe$^{3+}$
RI, the MPI and TI to Xe$^+$ {\it and} the TI to Xe$^{2+}$ is
fitted as there is more than one recollision mechanism possible
resulting in triple ionization.

By comparing the quality of fit of MPI, TI, FI and RI to the data
in Fig. 4, we can directly determine how successfully the
geometry- and diffraction-independent PPIs are predicted by
theory. As the PPI measured with circular polarization is not
influenced by RI, the circular data is therefore a more direct
comparison to the predicted MPI and TI response.

For ionization to Xe$^+$, MPI with the absorption of eight photons
occurs at the lowest intensities ($<$ 2.5 $\times$ 10$^{13}$
Wcm$^{-2}$) present in Fig. 4, indicating that LOPT is applicable.
As the laser intensity increases, the PPI response tends away from
MPI: the TI prediction fits the data excellently (i.e. directly
predicted by Keldysh theory), even at intensities greater than
that where FI is predicted to be the dominant mechanism. The
Xe$^+$ data clearly illustrates the intensity ranges over which
the different ionization mechanisms apply.

When the laser field generates Xe$^{2+}$, the absorption of
fourteen photons is required for MPI to proceed. As is apparent
from the long dashed line, only the very lowest intensities are
predicted by MPI. TI theory is even more successful in this,
accurately predicting a three order of magnitude increase in the
PPI to better than 10 \%. However, it is necessary to translate
the data by a factor of 0.91 in intensity to achieve this fit (cf
0.85 from \cite{yama}, albeit during a shorter duration laser
pulse, and geometry-dependent ion yield was measured). As with
$\it n$ = 1, FI is only reasonably accurate. In the case of the
linearly polarized laser pulse, at low intensity, there is
considerable RI present, as indicated by the short dashed line.
The RI contribution follows clearly the shape of the theoretical
prediction of sequential MPI and TI production (short dashed line)
visually fitted to the Xe$^{2+}$ data. This is consistent with
Xe$^+$ being the source atom for RI to Xe$^{2+}$. Such a method
has been used by a number of groups (for example, see \cite{laro})
fitting RI in geometry-dependent ion yield measurements, and it
appears to be successful here.

LOPT does not apply at all in the case of Xe$^{3+}$ as the
expected gradient of 20.6 is far too high to be supported by the
data. However, TI theory is able to accurately predict the PPI
from low intensity right up to saturation, with the data
translated by a factor of 0.85 in intensity, (cf 0.70 from
\cite{yama}), here well-defined by FI (vertical arrow). There are
now three possible nonsequential RI routes possible: (a) 0
$\rightarrow$ 1 $\Rightarrow$ 3, (b) 0 $\Rightarrow$ 2
$\rightarrow$ 3 and (c) 0 $\Rrightarrow$ 3, where $\rightarrow$
indicates sequential ionization, $\Rightarrow$ indicates double
ionization through RI, and $\Rrightarrow$ indicates triple
ionization through RI. While there is strong evidence for
mechanisms (a) and (b) as is clear from the two short dashed lines
on Fig. 4, there is little evidence for the presence of mechanism
(c). If it does occur, it is with a greatly suppressed PPI as
compared to (a) and (b). The shape of the RI PPI for {\it n} = 3
is well described by a combination of the theoretical prediction
of sequential ionization to Xe$^+$ and Xe$^{2+}$ as denoted by the
short dashed lines (a) and (b) in Fig. 4 respectively.

The ease with which ionization mechanism may be determined from
Fig.~\ref{fig4} is illustrated by considering what happens to a
group of atoms in a circularly polarized laser field at an
intensity of 2 $\times$ 10$^{14}$ Wcm$^{-2}$. Such an intensity
will at least triply ionize the atom: all atoms will be ionized to
Xe$^+$ by FI, between 40\% and 50\% of these ions will be further
ionized to Xe$^{2+}$ by TI, and of these ions, 1\% will undergo
MPI to Xe$^{3+}$. The situation with linear polarization is
complicated by NSMI, however the contribution is readily
calculable from Fig.~\ref{fig4}. Through a combination of
multiphoton, tunnel and field ionization theory, the response of
xenon to the ultrafast laser field may be accurately quantified.

\subsection{Ionization mechanism\\and the Keldysh
parameter} The pioneering theoretical work of Keldysh \cite{keld}
established a rule for distinguishing between ionization
mechanisms in a laser field using the fact that the ionization is
governed by the relative frequency of the laser field and the
tunnel frequency of the electron. The ratio of these frequencies
is defined as the adiabaticity, or Keldysh parameter, $\gamma$
\cite{keld, ppt}, and allows the ionization mechanism to be
broadly determined:
\begin{eqnarray}
\nonumber
\gamma &=& \frac{\omega_{laser}}{\omega_{tunnel}} =
{\sqrt{\frac{E_i}{2U_p}}}\hspace{20pt}{\rm where}\\
\nonumber
\gamma&\gg&1\hspace{90pt}{\rm MPI}\\
\nonumber
\gamma&\simeq&1\hspace{90pt}{\rm TI}\\
\gamma&\ll&1\hspace{90pt}{\rm FI}
\end{eqnarray}
In Eq. (3), {\it U}$_p$ = 9.33 $\times$ 10$^{-14}$ {\it I}
$\lambda^2$ is the pondermotive potential of the laser field, with
intensity {\it I} in Wcm$^{-2}$ and laser wavelength $\lambda$ in
$\mu$m, and {\it E}$_i$ is the ionization potential of the atom in
eV. The accepted definition is that when $\gamma$ $\gg$ 1, the
frequency of the laser is greater than the tunnelling frequency,
hence MPI results. When $\gamma$ $\simeq$ 1, the frequency of the
laser field is comparable to the tunnel frequency of the electron,
hence tunnel ionization is the most prominent mechanism.
Conversely, when $\gamma$ $\ll$ 1, the laser field comparable to
the Coulomb field between the nucleus and the electron, therefore
classical FI dominates. To determine how well the Keldysh
adiabaticity parameter quantifies ultrafast ionization of xenon,
our quantification of the ionization mechanism as presented in
Fig. 4 is converted to adiabaticity in Fig. 5, and the data
normalized to unity at $\gamma$ = 0.1.
\begin{figure}
\includegraphics[width=235pt]{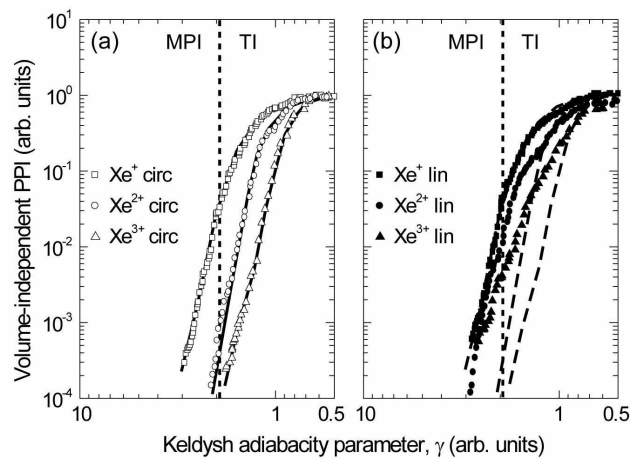}\\
\caption{PPI as a function of Keldysh parameter for Xe$^{n+}$
({\it n} = 1, 2, 3) for (a) circular and (b) linear polarization.
The presence of multiphoton ionization (MPI) dominates at high
$\gamma$, whereas tunnel ionization (TI) is more prominent at
$\gamma$ $<$ 2. A five point running average (solid line) through
the points in (a) is duplicated onto (b) (dashed line) to
illustrate the sequential ionization PPI as a function of
$\gamma$. Interestingly, the contribution from recollision
ionization in the linear pulse tends to MPI/TI response of
Xe$^+$.}\label{fig5}
\end{figure}

In Fig.~\ref{fig5}(a), the PPI to Xe$^{\it n+}$ ({\it n} = 1, 2,
3) in a circularly polarized laser pulse as a function of $\gamma$
is presented. Interestingly, the transition between MPI and TI for
$\it n$ = 1 and $\it n$ = 2 occurs at the same $\gamma$ = 2, thus
the identification of the presence of MPI at low intensities in
Xe$^{2+}$ is realistic. The vertical dashed line indicates the
transition between MPI and TI occurring at $\gamma$ = 2, which is
consistent with the intensities at which the transition between
MPI and TI is observed for Xe$^+$ and Xe$^{2+}$ (see Fig. 4). The
position of this transition is consistent with the prediction of
Keldysh. Furthermore, the PPI of Xe$^{\it n+}$ ({\it n} = 1, 2, 3)
all tend to unity at $\gamma$ = 0.5. For $\gamma$ $<$ 0.5, FI
dominates for ({\it n} = 1, 2, 3). Fig. 5(a) also contains a
5-point running average through the experimental points (solid
lines), to indicate the trend of the data. These trends are
duplicated onto Fig. 5(b) as dashed lines, allowing the comparison
of PPI in circular and linear polarized laser pulses.

As is apparent from Fig.~\ref{fig5}(b), the PPI of Xe$^{\it n+}$
({\it n} = 1, 2, 3) in a linearly polarized laser pulse as a
function of $\gamma$ is rather different from the circular case,
as expected from Fig. 3 and 4 due to the presence of RI. When $\it
n$ = 1, there is little deviation between linear and circular over
the full range of $\gamma$ recorded. However, for $\it n$ = 2,
around $\gamma$ = 1.2, the PPI breaks away from the response
observed in the circular case (long dashed line), and tends to
follow the PPI response for $\it n$ = 1. This transition, due to
RI, has not been observed in this manner before. Importantly, the
$\it n$ = 1 PPI is the upper limit of PPI enhancement by RI for
$\it n$ = 2. As is apparent from Fig. 5(b), the PPI as a function
of $\gamma$ for $\it n$ = 3 exhibits a similar behaviour, however
as there are now two RI processes present, as $\gamma$ is
decreased, the PPI for {\it n} = 3 tends first to the PPI for $\it
n$ = 2 around $\gamma$ = 1.3, then around $\gamma$ = 2.3, tends to
the $\it n$ = 1 response.

\section{Summary}
A new investigation into atomic ionization dynamics has been
presented, employing a novel experimental technique allowing a
direct comparison between linear and circular polarizations. The
key to effective intensity matching (EIM) is defining the spatial
dependence of ionization yield as independent of polarization
type. Intensity selective scanning (ISS) is used to measure the
ionization of xenon as a function of laser intensity by
translating the focusing optic with respect to a spatially limited
time-of-flight spectrometer.

By deriving a solution for the diffraction of a Gaussian laser
pulse through an arbitrary ABCD optical system, geometric effects
have been removed from the EIM-ISS results, producing partial
probabilities of ionization (PPI) for Xe$^{\it n+}$ ({\it n} = 1,
2, 3) for both linear and circular polarizations. This technique
has allowed clear measurement of the PPI due to recollision
ionization, which contributes significantly to double and triple
PPI in a linear polarized laser field. Multiphoton, tunnel and
field ionization contributions are clearly identified for the
charge states presented for both linear and circular
polarizations. We find that for the lowest intensities,
multiphoton ionization successfully predicts ionization to Xe$^+$
and Xe$^{2+}$. As the laser intensity is increased, tunnel
ionization theory is extremely successful, up to the intensity at
which classical field ionization dominates.

The volume-independent PPIs have also been presented as a function
of the Keldysh adiabaticity parameter for linear and circular
polarizations. The applicability of multiphoton and tunnel
ionization theory allows a precise definition of the transition
between ionization mechanisms, which occurs when the Keldysh
adiabaticity parameter, $\gamma$ = 2. Furthermore, an interesting
dependence on the lower ionization states is observed for
recollision ionization in a linearly polarized laser pulse. This
due to the necessity to generate the source ion before recollision
can proceed.

\begin{acknowledgments}
This research was supported by the Engineering and Physical
Sciences Research Council (EPSRC), UK: both EMLE and SLW
acknowledge studentships from the EPSRC, JMK would also like to
acknowledge funding from DEL, MS from IRCEP at QUB.
\end{acknowledgments}

\appendix
\section{Diffraction quantification}
The conventional approach for treating the propagation of light
beams is through the solution of the generalized Huygens-Fresnel
integral (see for example \cite{bornwolf}). Specifically, the
Collins form \cite{collins} of this integral is often used, as it
allows the ABCD system matrix to be directly incorporated into the
calculation. When the propagation of the beam is limited by some
form of aperture, solving the Huygens-Fresnel integral
analytically becomes more difficult, and it often requires a
numerical solution. The difficulty then becomes one of generating
a sufficiently accurate, computationally efficient solution. This
has been addressed recently by L\"{u} and Ji \cite{luji}, who
deftly adapted an approximation of the aperture function of Wen
and Breazeale \cite{wen} to allow the analytical treatment of
multi-apertured ABCD optical systems for the first time. However,
the approximate aperture function employed introduces oscillations
into the light amplitude even before any diffraction effects are
present, thus it is felt that a more accurate solution to the
problem of propagating Gaussian beams through apertured ABCD
optical systems is warranted.

\subsection{General solution}
In considering the Huygens-Fresnel integral \cite{bornwolf} for an
arbitrary ABCD system matrix in polar co-ordinates, the beam
intensity on the output plane {\it U}\hspace{2pt}({\it r}$_{2}$,
$\phi$, {\it z}) is defined as
\begin{equation}
U(r_2, \phi, z) = \frac{1}{\lambda B}\hspace{-4pt}\int_0^{2\pi}
\hspace{-8pt}\int_0^{\infty} \hspace{-4pt}U_0(r_1, \theta,
z\hspace{-2pt}=0) \exp({\rm i}k
S)\hspace{2pt}r_1\hspace{2pt}dr_1\hspace{2pt} d\theta\
\end{equation}
also referred to as the Collins diffraction integral
\cite{collins}, where
\begin{equation}
S = z + \frac{1}{2B}(A r_1^2 - 2r_1 r_2 \cos[\theta - \phi] + D
r_2^2)
\end{equation}
is the path between point ($r_1$, $\theta$) on the input plane and
point ($r_2$, $\phi$) on the output plane, and \textit{z} is the
distance between the input and output planes. Variables
\textit{A}, \textit{B} and \textit{D} are the elements of the
system matrix. For a Gaussian input beam, in the plane defined by
{\it r}$_1$ and {\it z} = 0, we define
\begin{equation}
U_0(r_1, \theta, z\hspace{-2pt}=0) = \exp ({-r_1^2}/{\omega_g^2})
\end{equation}
where $\omega_g$ defines the beam radius in the input plane. By
substituting Eq. (A3) and Eq. (A2) into Eq. (A1), and employing
the following expression from laser resonator calculations
\cite{lasres}
\begin{eqnarray}
\int_0^{2\pi}\exp\left[{\rm i}\left(\frac{k r_1 r_2}{X}\cos(\theta
-\phi) - l\theta\right)\right]d\theta\nonumber\\
= 2\pi {\rm i}^l\exp(-{\rm i}\hspace{2pt}l \phi) J_l \left(\frac{k
r_1 r_2}{X}\right)
\end{eqnarray}
where $\textit{J}_{l}$ is a Bessel function of the first kind and
$\textit{l}\hspace{3pt}^{\rm th}$ order, we arrive at the
following general expression for $\textit{l}$ = 0:
\begin{eqnarray}
U(r_2, \phi, z) = \frac{2\pi}{\lambda B}\exp \left(\frac{{\rm
i}kDr_2^2}{2B}\right)\exp ({{\rm i}kz})\nonumber\\
\hspace{-0pt}\times \int_0^{\infty}\exp(P r_1^2) ~J_0 (Q r_1)
\hspace{2pt}r_1 \hspace{2pt}dr_1\nonumber
\end{eqnarray}
where
\begin{eqnarray}
P &=& -\frac{1}{\omega_g^2} + \frac{{\rm i}\hspace{2pt} k A}{2B}\nonumber\\
Q &=& \frac{k r_2}{B}
\end{eqnarray}
and {\it k} = 2$\pi$/$\lambda$. The analytical solution of this
integral is now examined for the unapertured and apertured cases.
It is assumed that the ABCD optical system generates a focus in
the vicinity of the output plane, however the solutions presented
apply for $\it{any}$ system matrix.

\subsection{Unapertured solutions}
Many contemporary laser-dilute matter experiments assume an
unapertured Gaussian pulse profile, which when focused generates
the following well-known distribution of intensity, {\it
I}\hspace{2pt}(\hspace{-1pt}{\it
r}\hspace{2pt}$_2$,\hspace{2pt}{\it z})
\begin{equation}
I(r_2,z) = \frac{I_0}{1 + (z / z_0)^2}\exp\left(\frac{-2
r_2^2}{\omega_0^2\hspace{2pt}[I_0 / [1 + (z / z_0)^2]}\right)
\end{equation}
where {\it r}\hspace{2pt}$_2$ and {\it z} define the co-ordinate
frame in the vicinity of the focus, and the the beam waist,
$\omega_0$ = 2{\it f}\hspace{2pt}$\lambda$/$\pi${\it $\omega_g$}
and Rayleigh range, ${\it z}_0$ =
$\pi\hspace{2pt}\omega_0^2$/$\lambda$ characterize the intensity
distribution. In these expressions, \textit{f} is focal length of
the lens generating the focus, $\lambda$ the wavelength and {\it
$\omega_g$} the 1/{\it e}$^2$ unfocussed beam diameter. The on
axis distribution can be trivially found by setting {\it r}$_2$ =
0. Solving Eq. (A5) for {\it r}$_2$ = 0, the on-axis unapertured
solution for the current treatment is found by evaluating the
integral $\int_0^\infty$d{\it r}$_1$, giving
\begin{equation}
U(r_2 = 0,z) = -\frac{\pi}{\lambda B P}\exp({\rm i}kz)
\end{equation}
This expression may be expanded to generate the off-axis
unapertured solution, by performing the integration in Eq. (A5) as
before for $\int_0^\infty$d{\it r}$_1$, but with {\it r}$_2$
$\neq$ 0, with a solution
\begin{equation}
U(r_2, z) = -\frac{\pi}{\lambda BP}\exp\left(\frac{{\rm
i}kDr_2^2}{2B}\right)\exp({\rm i} kz)\exp(Q^2/4P)
\end{equation}
This solution requires the real part of {\it P} to be negative,
and the imaginary part of {\it Q} to be zero. As this is satisfied
for all cases by the definitions in Eq. (5), this is a universal
solution.

\subsection{Apertured solutions}
To allow for the action of a finite aperture before the ABCD
optical system, the integral in Eq. (A5) is evaluated
$\int_0^a$d{\it r}$_1$ where {\it a} is the aperture radius. The
on-axis solution is found by setting {\it r}$_2$ = 0, producing
the following expression
\begin{equation}
U(r_2 = 0, z)=\frac{\pi}{\lambda BP}\exp({\rm i}kz)(\exp (a^2 P) -
1)
\end{equation}
where the conditions applied to Eq. (A5) are applicable here.

To find a solution to the off-axis apertured system, the integral
in Eq. (A5) is evaluated $\int_0^a$d{\it r}$_1$ where {\it a} is
the aperture radius. This equation cannot be solved analytically,
so two methods of calculating the spatial intensity distribution
are employed. Firstly, we perform a Taylor Series expansion around
{\it r}$_2$ = 0, of the integral term in Eq. (A5), so the general
form is thus
\begin{eqnarray}
U(r_2, z) &=& \frac{2\pi}{\lambda B}\exp \left[{\rm
i}k\left(\frac{Dr_2^2}{2B} + z\right)\right]\nonumber \\
&\times& \sum_{y = 0}^\infty \sum_{x = 0}^y\left(\frac{-1^y}{(y -
x)!}\frac{P^{y - x}\hspace{3pt}Q^{2x}}{2^{2x} (x!)^2}\right)
\end{eqnarray}
where {\it x} and {\it y} are the indices of the expansion terms.
This solution can be used to calculate the distribution of
intensity for certain optical systems depending on {\it P}, {\it
Q} and {\it a}. For a relatively sharp focus ({\it
f}$\hspace{4pt}$of the order of $\omega_g$) with little
diffraction ({\it a} $\gg$ $\omega_g$), the number of terms
required in the Taylor Series is small. However, when the system
exceeds these limits, particularly when there is significant
diffraction of the incoming beam, the number of terms required
becomes untenable, and the ({\it x}\hspace{2pt}!)$^2$ term
requires impracticable computational power to evaluate. In this
case, the intensity distribution may be evaluated by converting to
a finite element problem through
\begin{eqnarray}
U(r_2, z) &=& \frac{2\pi}{\lambda B}\exp \left[{\rm
i}k\left(\frac{Dr_2^2}{2B} + z\right)\right]\nonumber \\
&\times& \sum_{r_1 =0}^a r_1 \exp(P r_1^2) J_0(Q r_1) \Delta r_1
\end{eqnarray}
This solution requires accurate and rapid evaluation of the Bessel
Function, which becomes processor-intensive as {\it Qr}$_1$
becomes large. By approximating {\it J}$_n$({\it Qr}$_1$) $\simeq$
(2/$\pi${\it Qr}$_1$)$^{1/2}$ $\cos$[{\it Qr}$_1$ - ({\it
n}$\pi$/2) - ($\pi$/4)] when {\it Qr}$_1$ $>$ 8, evaluation of
this sum becomes very efficient while retaining a deviation from
the expected value of $<$ 1 in 10$^4$. The off-axis solution is
therefore
\begin{widetext}
\begin{equation}
U(r_2, z) = \frac{2\pi}{\lambda B}\exp \left[{\rm
i}k\left(\frac{Dr_2^2}{2B} + z\right)\right] \sum_{r_1 =0}^a r_1
\exp(P r_1^2)\left(\frac{2}{\pi r_1
Q}\right)^{1/2}\hspace{-4pt}\cos\left(Q r_1 -
\frac{\pi}{4}\right)\Delta r_1
\end{equation}
\end{widetext}
By making $\Delta$\hspace{0pt}{\it r}$_1$ small, Eq. (A11) and
(A12) will produce an extremely accurate quantification of the
output plane. The selection of which form to employ depends on the
geometry of the ABCD system and the size of the incoming beam
relative to the aperture diameter. Computational efficiency may be
further improved through the application of Adaptive Mesh
Refinement \cite{amr}, where the grid spacing depends on the rate
of variation in {\it U} resolved along both the {\it r}$_2$ and
{\it z} axes.

An important test for the off-axis apertured solution, Eq. (A10)
to Eq. (A12) is to check for convergence to the off-axis
unapertured solution, Eq. (A8) and the accepted definition, Eq.
(A6) when {\it a} $\gg \omega_g$. Through rigorous two-dimensional
comparisons for a variety of ABCD systems, beam radius $\omega_g$
and aperture radius {\it a} have shown that all solutions
presented are self-consistent, and, when applicable, consistent
with the accepted solution.

\end{document}